\documentclass[lettersize,journal]{IEEEtran}
\usepackage{amsmath,amsfonts}
\usepackage{algorithmic}
\usepackage{algorithm}
\usepackage{array}
\usepackage[caption=false,font=normalsize,labelfont=sf,textfont=sf]{subfig}
\usepackage{textcomp}
\usepackage{stfloats}
\usepackage{url}
\usepackage{pgfplots}
\usepackage{adjustbox}
\usepackage{tikz}
\usepackage{xcolor}
\usepackage{colortbl}
\usepackage{verbatim}
\usepackage{graphicx}
\usepackage{cite}
\begin{document}

\title{Pretext Tasks Selection for Multitask Self-Supervised Audio Representation Learning}


\author{
  Salah Zaiem, Télécom Paris, \textit{salah.zaiem@telecom-paris.fr} \\
   Titouan Parcollet, LIA, Avignon Université, \textit{titouan.parcollet@univ-avignon.fr}   \\
  Slim Essid, Télécom Paris, \textit{slim.essid@telecom-paris.fr} \\
  Abdelwahab Heba, Université Paul Sabatier, IRIT, CNRS,
\textit{aheba@irit.fr}
}
\markboth{Preprint}%
{Shell \MakeLowercase{\textit{et al.}}: A Sample Article Using IEEEtran.cls for IEEE Journals}


\maketitle

\begin{abstract}
Through solving pretext tasks, self-supervised learning leverages unlabeled data to extract useful latent representations replacing traditional input features in the downstream task. In audio and speech signal processing, a wide range of features were engineered through decades of research efforts. As it turns out, learning to predict such features  has proven to be a particularly relevant pretext task, leading to useful self-supervised representations which prove to be effective for downstream tasks. However, methods and common practices for combining such pretext tasks for better performance on the downstream task have not been explored and understood properly. In fact, the process relies almost exclusively on a computationally heavy experimental procedure, which becomes intractable with the increase of the number of pretext tasks. This paper introduces a method to select a group of pretext tasks among a set of candidates. The method we propose estimates calibrated weights for the partial losses corresponding to the considered pretext tasks during the self-supervised training process. The experiments conducted on automatic speech recognition, speaker and emotion recognition and instrument classification validate our approach as the groups selected and weighted with our method perform better than classic baselines, thus facilitating the selection and combination of relevant pretext-task labels for self-supervised representation learning.
\end{abstract}


\section{Introduction}

Self-supervised learning (SSL) methods usually rely on a supervision obtained from the data itself through solving specific pretext tasks leveraging the underlying structure of the considered data \cite{doersch2016unsupervised, Arandjelovic_2018_ECCV}. This technique is used in various domains including image processing \cite{misra2020self, jing2020self, grill2020bootstrap}, natural language understanding \cite{chen2020big, du, lan2019albert} or speech and audio processing \cite{baevski2020wav2vec,Liu_2020,jiang2020speech}. It offers numerous advantages, such as the independence from labeled data, stronger performance on downstream tasks, more robust models and an easier transfer to low-resource setups (\textit{e.g.}, low-resource languages) \cite{baevski2020wav2vec, jing2020self}.   

The numerous existing SSL approaches are characterized by the nature of the pretext tasks they solve. For instance, common techniques include predictive coding \cite{baevski2020wav2vec,Liu_2020,song2020speechxlnet,Zhang2020, hubert}, pretext-task label learning \cite{pascual2019learning,ravanelli2020multitask}, auto-encoding \cite{Renshaw2015ACO,algayres}, triplet-loss learning \cite{shor2020towards,peplinski2020frill}, generative modelling \cite{khurana2020convolutional}, contrastive learning \cite{Saeed2020,jiang2020speech}, denoised masked speech modelling \cite{chen2021wavlm}, and leveraging labeled and unlabeled data in hybrid trainings \cite{wang2021unispeech}. More precisely, these pretext tasks may be defined through the choice of pretext labels, hereafter referred to as \textit{pretext-task labels}. The automatic extraction of pretext-task labels for SSL (\textit{i.e.} from the data itself) is common in many application domains, such as computer vision \cite{noroozi2017unsupervised, gidaris}, music processing \cite{hung2019multitask, wu2021multitask}, speech processing \cite{pascual2019learning, shukla} and is commonly referred to as \textit{multitask self supervised learning}. In the specific context of speech processing, the process of designing pretext-task labels may benefit from decades of research in signal processing. For instance, potential candidates are pitch estimations, energy-based features, voicing state, noise estimations and many more.

As demonstrated by \cite{pascual2019learning, wu2021multitask}, multitask speech representation learning is a powerful tool to build representations that are beneficial for a wide range of distinct downstream tasks by combining different pretext-task labels which intuitively correspond to these tasks. Unfortunately, there is no clear understanding of the pretext-task labels' complementarity during joint multi-task SSL training, and therefore, no common practice describing a group selection strategy for pretext-task labels to obtain better performance on a known downstream task. As a matter of fact, this design process has been essentially driven by empirical validation. This empirical approach can rapidly become intractable with modern SSL architectures which may contain billions of parameters trained on thousands of hours of speech, not to mention the carbon footprint of such pretext-task label searches. For instance, the self-supervised training of a single state-of-the-art large wav2vec 2.0 model \cite{baevski2020wav2vec} on $53.2k$ hours of speech currently requires $128$ GPUs for $5.2$ days.

This paper builds on previously published work \cite{zaiem2021conditional} which validated Conditional Independence (CI) for individual pretext-task selection. It aims at providing a clear, efficient and theoretically motivated procedure for  pretext-task label group selection and weighting based on CI. The method presented allows one to design ahead of training the most adapted multitask self-supervised speech representation learning model which perfectly suits the considered downstream tasks. Such an approach may also enable researchers to save a substantial amount of time and compute  devoted to pretext-task label search. Hence, the contributions of this work are fourfold:   

\begin{enumerate}   
    \item Introduce a theoretically motivated and computationally efficient method for the selection of \textit{groups} of pretext-task label among a set of candidates and with respect to the considered downstream tasks (Sections \ref{sec:CI} and \ref{sec:group}).
    \item Validate empirically the proposed approach with a first SSL model relying on different sets of pretext-task labels, corresponding to the ones obtained for three considered speech tasks. (Sections \ref{sec:exps}). 
    \item Extend our method to state-of-the-art architectures such as wav2vec 2.0 to enhance its performance and expose the scaling capabilities of our solution (Section \ref{sec:wav2vec}). 
    \item Perform a thorough study of the robustness and generalization potential of this technique to various changes including: type of data, pretraining and finetuning datasets and pretext-task candidates with an application on instrument classification. 
    \item Release the code base developed with SpeechBrain \cite{speechbrain} for replication and to encourage further investigations.\footnote{\url{https://github.com/salah-zaiem/Multitask-pretext-task-selection}}
\end{enumerate}
The conducted experiments demonstrate that the proposed method allows for a more intelligent, \textit{i.e.} better informed, pretext-task label group selection for multitask SSL settings. Indeed, we find that the models built with the proposed method obtain a word error rate, an equal error rate and an emotion recognition accuracy respectively, $31.6\%$ and $27.4\%$ and $7.8\%$ lower than the baseline, without the need for any empirical search on the pretext-task selection or weighting. When changing the pretraining dataset and the type of data to tackle musical instrument recognition, the accuracy of the best models is $11\%$ better than the baseline on solos and $4\%$ better in multi-instrument settings.
\section{Related works and motivations}
\label{sec:related}

 \begin{figure*}[t!]
  \centering
  \includegraphics[width=1\linewidth, scale=0.17]{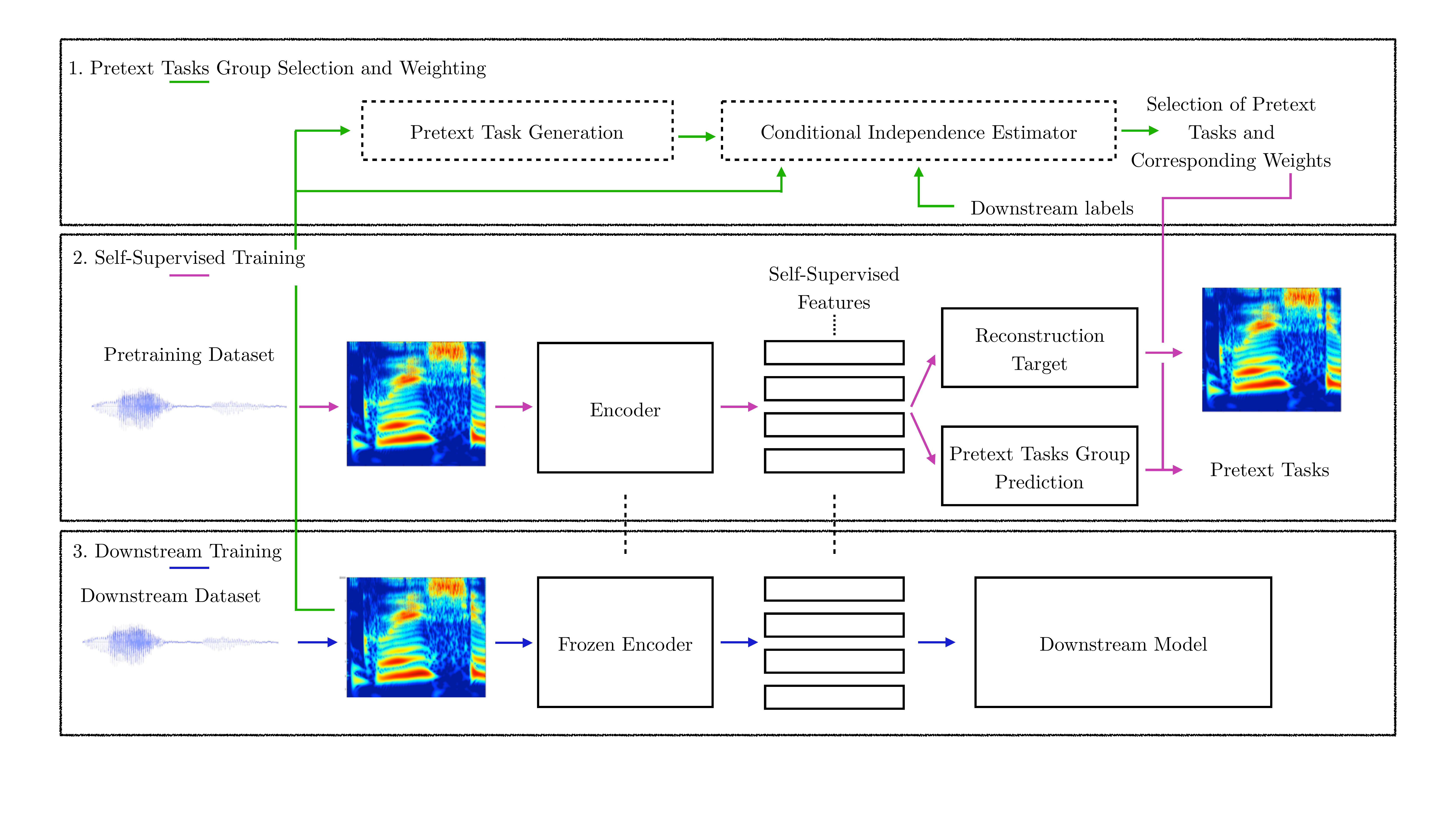}
  \vspace{-1.2cm}
  \caption{Illustration of the training pipeline. The three steps are depicted: 1. Selecting the group of pretext-task labels and their corresponding weights; 2. SSL training with the selected pretext task; 3. Training on the downstream task with the pretrained SSL model. }
  \label{diagram}
\end{figure*}

SSL recently became a key component to achieve good performance on downstream tasks especially with low-resource setups, either in speech \cite{baevski2020wav2vec, conneau}, natural language processing \cite{lan2019albert, chen2020big} or computer vision \cite{gidaris2019boosting, misra2020self, jing2020self}. Due to its very nature, SSL relies on large amounts of unlabeled data used to train large deep neural networks over long periods of time. It is thus crucial to understand properly what makes for a good pretext task for a SSL model to lower the amount of computation and time needed to obtain the best downstream performance.

\textbf{SSL for Speech.}  Self-supervised learning for speech has recently enabled researchers to reach state-of-the-art results on various speech processing tasks \cite{fan2021exploring}. The most successful models rely on predictive and contrastive objectives  \cite{baevski2020wav2vec, chung2019unsupervised, hubert,  shor2021universal} performing well across the different tasks even in low-resource settings. This led to the design of different benchmarks evaluating the self-supervised representations in different languages \cite{superb, evain2021lebenchmark}. However, in contrast to this proposition,  these works have not tried to  theoretically motivate beforehand the pretext task choices made in the self-supervision pipeline.

\textbf{Understanding SSL.} A few works have tried to shed some theoretical light on the mainly empirical field of self-supervised learning. Following the different paradigms in SSL, various tracks have been followed to understand what makes for a good self-supervised representation, exploring different approaches \cite{ lee2020predicting, Arora2019, wei}. On the one hand, contrastive learning \cite{oord2018representation, chen2020simple, xiao2021contrastive} has been advocated both theoretically and empirically to achieve a balance in the mutual information between alternative representations of the data, keeping just enough shared information to retain the class-related content \cite{tschannen2019mutual, tian, bachman}. In a recent work \cite{li2021selfsupervised}, independence testing has been used to produce better transformations in contrastive learning settings for image representations. Predictive learning, on the other hand, requires the model to predict masked elements in sequential data. This technique is powerful on downstream tasks that can be reduced to a masking problem, as suggested by research on language modeling \cite{saunchi}. However, all these works have been focusing solely on computer vision or text-related applications, and none of them addressed the multi-task self supervision problem.

\textbf{Multi-task self-supervised learning.} While the literature on multi-tasking in self-supervised learning remains scarce, it has been shown in classic supervised learning settings, that through estimates of similarity between tasks or thorough empirical testing, several tasks can take advantage of being solved with a common encoder \cite{zamir2018, Dwivedi2019, shafey, Chen2015}. More specifically, combining pretext tasks with SSL has been mainly explored in computer vision and speech \cite{pascual2019learning, ravanelli2020multitask}. Pretext tasks such as Jigsaw \cite{doersch2016unsupervised}, colourisation and rotation \cite{gidaris} have been combined successfully to improve downstream performance \cite{jigsawpluss, ShinyaYamaguchi}. The two closest works to our line of research are from Lee et al. \cite{lee2020predicting} and Doersch et al. \cite{Doersch}. The former shows that a theoretical link can be established between the conditional independence of the data points and their pretext-task value given the downstream label, and an improvement of the performance on the downstream task, while the latter proposes to select layers from a multitask self-supervised encoder according to the pretext task to be solved. However, in both cases, while succeeding in presenting a proof-of-concept in multitask speech SSL training, the studies do not offer practical and theoretical solutions to select groups of pretext-task labels to build an adapted SSL model that will perform well on the considered downstream tasks. 

\textbf{Group feature selection.} Finally, feature selection, and especially group feature selection is another close and inspiring field given the problem we consider. The relationship and interactions between features have been largely investigated in the supervised learning literature \cite{guyon}. This led to multiple solutions to the feature group selection problem, including LASSO-based techniques \cite{grouplasso}, or multiple kernel formulations \cite{sonnenburg, bach}. Another type of popular solutions came from the research on submodularity, leading to information-theoretically motivated group selections \cite{9592798,pmlr-v37-wei15}. This has been tried specifically on speech to avoid domain mismatch harming the final downstream performance \cite{mortaza}. Especially on speech,  However, these works do not involve any self-supervision, and links between feature selection, self-supervision design and pretext-task selection are yet to be proved. In the experiments section (Section \ref{sec:exps}), we will consider these lines of works as concurrent baselines.

With this work, we aim at shortening the process of designing SSL models through giving insights on how to select suitable pretext tasks towards solving a given dowstream one. We decided to experiment primarily with audio data due to the lack of literature on this domain for multitask SSL, and for the various pretext-task labels available, which are based on decades of signal processing research, before extending to music data. The whole pipeline starting from the acoustic feature extraction to the downstream task scoring follows three major steps summarized in Figure \ref{diagram}. First, for every downstream task, our method produces a pretext task selection and weighting. Then, a SSL model is trained, before being used as a feature extractor front-end to one or many downstream tasks. 

\section{Conditional independence for utility estimation}
\label{sec:CI}

As a first step, we require a function that estimates the utility of learning to solve a pretext-task to improve the performance on the downstream task. We use an estimation of the conditional independence between the pretext-task label values and the downstream data points given the downstream labels. Hereafter, we explain the theoretical motivations and describe the computation steps. 

\subsection{Problem definition and intuition}

 Let $X$, $Y$ and $Z$ be, respectively, the downstream data points, their downstream labels and their pretext-task labels. Let also $\mathcal{C}$ be the set of possible downstream classes. As an example, if one considers speaker recognition as a downstream task, $X$ would be the speech samples, $Y$ the speaker IDs, $\mathcal{C}$ the set of unique speaker IDs, and $Z$ an automatically computed signal feature, such as the fundamental frequency.

 As stated in Section~\ref{sec:related}, Lee et al. \cite{lee2020predicting} linked the utility of a pretext task defined by the prediction of a pretext-task label ($Z$) to the conditional independence (CI) between $Z$ and $X$ given $Y$. The approach prescribes that, given the labels $Y$, one may seek to \textit{quantify how much it is possible to predict the pretext-task labels $Z$ without knowing much about $X$}. The authors bounded, under certain assumptions, the downstream classifier error with a function of the downstream training set size, and a measure of the CI. More precisely, the main theorem shows that the bounding function decreases linearly with the downstream-task dataset size ($M$) and quadratically with the CI, which indicates a potential estimator for the pretext task utility. 
 
These results rely on two assumptions that are not upheld in the remaining of this work. First, the modelling functions are expected to be linear. Given the complexity of the considered downstream tasks, such as speech and speaker recognition, limiting ourselves to linear modelling would lead to very limited downstream performances.  Second, we will estimate the conditional independence using a kernelized independence test, while the quantity involved in the proven bounds is $ \epsilon^2_{CI} = \mathbb{E} [||\mathbb{E}[Z|X] - \mathbb{E}_Y [\mathbb{E}[Z|Y]|X]||^2]$. Computing this quantity is unpractical, especially with varying length speech samples while the method we chose to go with has been successfully tested on sequential data \cite{gretton}.
 
  What mainly holds is the intuition behind the use of conditional independence as a pretext task utility estimator. To get an intuitive understanding of the motivations of this choice, let us consider the example of image classification as the downstream task, and image colourization as the pretext task. In this case, this pretext task would be suited to the downstream one if the final classification label can help implying the colours. For instance, if there are only two classes "Blue skies" and "Yellow deserts", then colourisation is an interesting pretext task, as knowing the final label helps a lot for the pretext task, independently of the image. Similarly, if all the classes share the same colour palette, colourization may not be an interesting task. (In this toy example, we are ignoring the edge detection aspect of colourization, and only focusing on the colour choice part. Obviously, the former aspect plays a role in the success of the colourization pretext task)

 The proposed function depends on the final downstream task to be solved. This is motivated by two main reasons. First, it can be seen through the large literature on feature selection for various speech or computer vision tasks \cite{Liu_2020, serizel, schuller, Wang_2019_CVPR, loweimi}, that different tasks require the description of different aspects of the data.  This suggests that different downstream tasks may perform better after different pre-trainings. A second argument is the difficulty to evaluate representations quality intrinsically, \textit{i.e.} independently from the choice of a particular downstream task. A few metrics and tests \cite{schatz, carlin, lakhotia2021generative} have been proposed for speech, but the correlation between these and downstream-task performance has not been clearly identified \cite{algayres,Gump2020}. Finally, recent experiments adapting the self-supervised representation to the speaker identification task have shown substantial improvements compared to task-agnostic representations \cite{chen2021unispeechsat}, validating our intuition that downstream task oriented SSL is an interesting trend towards better downstream performances. We could also mention research in semi-supervised learning that managed to reach results comparable to the best SSL models through leveraging unlabeled data \cite{kaizen, https://doi.org/10.48550/arxiv.2110.00165}.

 The main issue with CI is the difficulty of computing an estimate of how much two variables are independent given a third one on realistic data \cite{hardness}. We will start with proposing a simple way to get an estimation of the conditional independence and validate it on individual pretext task selection.

\subsection{Conditional independence estimate computation}\label{method}

This section details the computation of the conditional independence estimate that is  used as a measure of pretext-task label utility. Let  $X =\{x_i\}_{i \in \{0, \ldots, M\}}$ with  with $x_i$ being data samples (\textit{e.g.}, Mel-band spectrogram for audio and speech, every $x_i$ here being the Mel-band spectrogram of a given speech segment). Every sample $x_i$ has a corresponding downstream label $y_i$ and an automatically generated pretext-task label $z_i$. We assume that $y_i$ is discrete reducing the task to a classification problem such as with speaker ID for speaker recognition. We also assume that for every pretext-task $Z$, a single $z_i$ value corresponds to each $x_i$. In our case, $z_i$ values are the mean of the frame-wise pretext-task label values.
 
For independence testing,  kernel-based Hilbert Schmidt Independence Criterion (HSIC) \cite{gretton} is used for two reasons. First, HSIC has already proven successful for textual data in testing statistical dependence between translated sentences \cite{gretton}. Second, kernel-based techniques facilitate the handling of multivariate and varying-length data such as speech, as the estimation then boils down to the computation of a similarity measure between the considered variables.

\textbf{Computation steps.} The estimation of the CI of a pretext-task label $Z$ for a downstream task $(X,Y)$ consists of three steps. We start by splitting the data samples $X$ according to the downstream (discrete) classes. Then, we compute for every downstream class $c \in \mathcal{C}$, the kernel matrices $K_c$ and $L_c$ representing the similarity measures for the data samples, and the pretext-task labels, respectively. Finally, we perform the independence test for every split group using $K_c$ and $L_c$ and aggregate the estimates with a weighted mean taking into account the number of samples per downstream class. Thus, for two speech samples $x_i$ and $x_j$, holding two pretext-task label values $z_i$ and $z_j$, the coefficients of the similarity matrices $K_c$ and $L_c$ are computed as follows:
\begin{align}
K_{ij} &= K(x_i,x_j) = \cos(GD(x_i) ,GD(x_j) ). \\
L_{ij} &= RBF(z_i, z_j),
\end{align}
with $GD(.)$ the Gaussian Downsampling function, $\cos(.,.)$ the cosine similarity, and $RBF(., .)$ the Radial Basis Function kernel, defined as: 
\begin{align}
    \cos(x,x') &= \frac{trace(x^T x')}{||x|| . ||x'||} . \\
    RBF(z,z') &= \exp (-\frac{|| z-z'||^2}{2 \sigma^2}),
\end{align}
where $\sigma$ is the width of the RBF kernel and $trace(.)$ the sum of elements of the main diagonal. Note that we compute the matrices $K_c$ and $L_c$, for each group of samples sharing the same downstream class $c \in C$. Hence, $K_c$ and $L_c$ correspond to the definitions above, but restricted to the points with $c$ as a downstream label. For each downstream class $c$, and as in \cite{gretton}, with $n_c$ being the number of points of class $c$, the HSIC value is given by:
\begin{equation}
HSIC_c(X, Z) = \frac{1}{n_c^2} trace(K_c H_c L_c H_c), 
\end{equation}
with $H_c= I_{n_c} - \frac{1}{n_c}1_{n_c} 1_{n_c}^T $, $n_c$ being the number of points with  label $c$, and $1_{n_c}$ a vector of ones of size $n_c \times 1$.

The HSIC value is non-negative and corresponds to the Hilbert norm of their cross-covariance matrix. It is used to characterize the independence of the two considered quantities. Intuitively, the HSIC value is high if samples similar in $K_c$ are similar in $L_c$. Therefore, the lower this value is, the more independent the two arguments of HSIC are and the better the pretext-task label should be for self-supervision before fine-tuning on the downstream class. The final value for a given pretext-task label and a downstream task is expressed as:
\begin{equation}
HSIC(X,Z |Y) = \frac{1}{M}\sum_{c \in \mathcal{C}} HSIC_c(X, Z) \times n_c.
\end{equation}
with $M$ being the number of points in the whole dataset.
\section{Validation on individual selection}\label{individual}
This section validates individual pretext task selection pretraining the encoder on the English Common Voice dataset and using the learned representations for two downstream tasks; Automatic Speech Recongition using TIMIT, and Speaker Verification using VoxCeleb1.

\textbf{SSL pretraining.} The train set of the English Common Voice dataset (version $5.1$) \cite{ardila2020common} is used for SSL pretraining ($700$ hours). Common Voice is a collection of speech utterances from worldwide users recording themselves from their own devices. Hence, the closeness to natural settings makes it a suitable choice for self-supervised learning. We remove from Common Voice the sentences lasting more than $10$ seconds, as they often contain long silence parts due to open microphones. 

\textbf{Downstream evaluation datasets.} TIMIT \cite{timit} is considered for the speech recognition task. It is composed of a standard $462$-speakers
training set, a $50$-speakers development set and a core test set of $192$ sentences for a total of $5$ hours of clean speech. For the CI estimation, and to get discrete labels to split on, we cut the sentences at the phone level, using the official transcripts. VoxCeleb1 \cite{Nagrani_2017} is used for the speaker verification task. The training set contains $148,642$ utterances from $1,251$ different speakers. The conditional independence is computed at the phone level for ASR and utterance level for speaker recognition making the assumption that phone segments are entirely independent samples

\textbf{Pretext-task labels and architecture details.} Based on previous work conclusions \cite{ravanelli2020multitask, jiang2020speech}, apart from the pretext-task label to be tested, our self-supervised model learns to reconstruct the input Mel spectrograms, and to compute $40$-dimensioned MFCC feature vectors. These targets are kept to avoid information loss harming heavily downstream performances. Inspired by the PASE model \cite{ravanelli2020multitask, pascual2019learning}, the model consists of an encoder followed by small predictors limited in capacity (more details on the architectures in appendix \ref{ap:details}) .

\begin{figure}[!t] 
\centering
\begin{adjustbox}{width=\linewidth}

\begin{tikzpicture}
\pgfplotsset{set layers}

\begin{axis}[
    width=\linewidth, height=7cm, 
    scale only axis,
    grid = major,
    grid style={dashed, gray!30},
    xmin=0.5,   
    xmax=7.5,  
    ymin=-0.25,   
    ymax=4.5,  
    axis background/.style={fill=white},
    axis y line*=right,
    tick align=outside,
     xticklabels={F0,Voicing,logHNR,specRastaL1, Loudness,PCM ZCR,Alpharatio},xtick={1,2,3,4,5,6,7},
  x tick label style={rotate=60,anchor=east}]
    mark repeat={600},
    ylabel near ticks,
    xlabel near ticks,
    xlabel style={text width=5cm},
    xlabel style={align=center},
    xlabel={test},
    legend pos=north west,
    legend columns=1,
    legend style={
            /tikz/column 2/.style={
                column sep=2pt,
            }
    }
    ]
\addplot+[mark=o, color=purple,line width=1pt] file {cit.txt};
\node [above=4 , text=black] at (axis cs:  1,  0.21) {$0.21$};
\node [below=5, text=black] at (axis cs:  2,  0.71) {$0.71$};
\node [above=4, text=black] at (axis cs:  3,  0.17) {$0.17$};
\node [above=4, text=black] at (axis cs:  4,  0.43) {$0.43$};
\node [above=4, text=black] at (axis cs:  5,  0.85) {$0.85$};
\node [above=4, text=black] at (axis cs:  6,  0.80) {$0.80$};
\node [above=5, text=black] at (axis cs:  7,  0.07) {$0.07$};

\addlegendentry{CI TIMIT}
\end{axis}
\begin{axis}[
    width=\linewidth, height=7cm, 
    scale only axis,
    grid = major,
    grid style={dashed, gray!30},
    xmin=0.5,   
    xmax=7.5,  
    ymin=15,   
    ymax=20,  
    axis background/.style={fill=white},
    axis y line*=left,
    axis x line=none,
    tick align=outside,
    xticklabels={F0,Voicing,logHNR,specRastaL1, Loudness,PCM ZCR,Alpharatio},xtick={1,2,3,4,5,6,7},
    ylabel near ticks,
    legend pos=north west,
    legend columns=1,
    legend style={
            /tikz/column 2/.style={
                column sep=2pt,
            }
    }    ] 

\addplot+[mark=o, color=orange,line width=1pt] file {PERTIMIT2.txt};
\node [above=4, text=black] at (axis cs:  1,  16.77) {$16.77$};
\node [above=4, text=black] at (axis cs:  2,  16.99) {$16.99$};
\node [above=8, text=black] at (axis cs:  3,  16.43) {$16.43$};
\node [above=4, left=1, text=black] at (axis cs:  4,  17.46) {$17.46$};
\node [above=3, text=black] at (axis cs:  5,  18.35) {$18.35$};
\node [above=4, text=black] at (axis cs:  6,  17.88) {$17.88$};
\node [above=15, text=black] at (axis cs:  7,  16.46) {$16.46$};

\addlegendentry{PER TIMIT}

\end{axis}

\end{tikzpicture}

\begin{tikzpicture}
\pgfplotsset{set layers}

\begin{axis}[
    width=\linewidth, height=7cm, 
    scale only axis,
    grid = major,
    grid style={dashed, gray!30},
    xmin=0.5,   
    xmax=7.5,  
    ymin=-0.25,   
    ymax=4.5,  
    axis background/.style={fill=white},
    axis y line*=right,
    tick align=outside,
     xticklabels={F0,Voicing,logHNR,specRastaL1, Loudness,PCM ZCR,Alpharatio},xtick={1,2,3,4,5,6,7},
  x tick label style={rotate=60,anchor=east}]
    mark repeat={600},
    ylabel near ticks,
    xlabel near ticks,
    xlabel style={text width=5cm},
    xlabel style={align=center},
    xlabel={test},
    legend pos=north west,
    legend columns=1,
    legend style={
            /tikz/column 2/.style={
                column sep=2pt,
            }
    }    ]
\addplot+[mark=o, color=purple,line width=1pt] file  {labeledcivc.txt};
\node [above=4 , text=black] at (axis cs:  1,  0.02) {$0.02$};
\node [below=5, text=black] at (axis cs:  2,  0.86) {$0.86$};
\node [above=4, text=black] at (axis cs:  3,  0.02) {$0.02$};
\node [below=4, text=black] at (axis cs:  4,  0.77) {$0.77$};
\node [below=3, text=black] at (axis cs:  5,  0.86) {$0.86$};
\node [below=4, text=black] at (axis cs:  6,  0.86) {$0.86$};
\node [above=5, text=black] at (axis cs:  7,  0.06) {$0.06$};

\addlegendentry{CI VC}
\end{axis}
\begin{axis}[
    width=\linewidth, height=7cm, 
    scale only axis,
    grid = major,
    grid style={dashed, gray!30},
    xmin=0.5,   
    xmax=7.5,  
    ymin=7.5,   
    ymax=15,  
    axis background/.style={fill=white},
    axis y line*=left,
    axis x line=none,
    tick align=outside,
    xticklabels={F0,Voicing,logHNR,specRastaL1, Loudness,PCM ZCR,Alpharatio},xtick={1,2,3,4,5,6,7},
    ylabel near ticks,
    legend pos=north west,
    legend columns=1,
    legend style={
            /tikz/column 2/.style={
                column sep=2pt,
            }
    }  ] 
\addplot+[mark=o, color=orange,line width=1pt] file {EERVCreal.txt};
\node [above=4, text=black] at (axis cs:  1,  9.99) {$9.99$};
\node [above=4, text=black] at (axis cs:  2,  9.98) {$9.98$};
\node [above=4, text=black] at (axis cs:  3,  9.08) {$9.08$};
\node [above=4, left=1, text=black] at (axis cs:  4,  9.32) {$9.32$};
\node [above=3, text=black] at (axis cs:  5,  12.68) {$12.68$};
\node [above=4, text=black] at (axis cs:  6,  10.1) {$10.1$};
\node [above=4, text=black] at (axis cs:  7,  10.01) {$10.01$};

\addlegendentry{EER VC}

\end{axis}

\end{tikzpicture}

\end{adjustbox}  
\caption {Left : Phone Error Rate and CI estimate values on TIMIT for every considered pretext-task label --- Right: Equal Error Rate and CI estimate values on VoxCeleb for every considered pretext-task label. Error rates appear on the left y axis. We can observe the monotonic relation between the estimator and the downstream errors, particularly for TIMIT.}
\label{timitresults}
\end{figure}
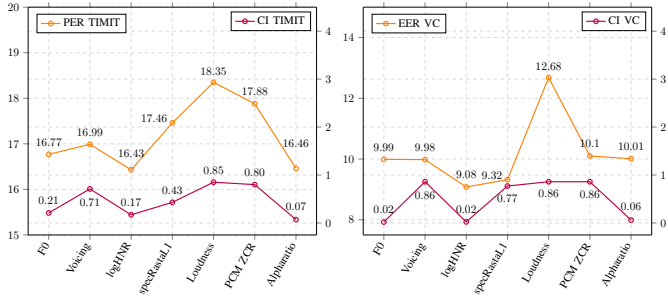

\textbf{Initial results.} Figure \ref{timitresults} summarizes the results of the experiment for all the considered pretext-task labels, reporting the CI estimates and the downstream performance for each of the two tasks. It shows the evolution of the conditional independence estimator and the PER and EER, respectively on TIMIT and VoxCeleb. In both figures, the two curves seem to follow the same trajectories showing a monotonic relationship.

According to theoretical insights\cite{lee2020predicting}, the lower the CI estimate is, the more independent is the pretext-task label from the  samples given the labels and the lower should the downstream error be. Hence, we are looking for a monotonic relationship between CI estimates and downstream errors. We consider two classic assessors of monotony: Spearman Correlation and Kendall $\tau$. While Pearson correlation measures the linear correlation between the values, Spearman correlation is a Pearson Correlation on the ranks of the values. Kendall $\tau$ considers all the pairs of pretext-task labels and checks whether their order in the CI estimate is the same for the error rate ( \textit{i.e} the pair is concordant ). The more concordant pairs there are, the higher Kendall $\tau$ is.

Spearman correlations reach 0.48 for speaker recognition and a high 0.93 on TIMIT for ASR, while Kendall $\tau$ is respectively 0.41 and 0.81 for the two tasks. The correlations between CI and the downstream error are logically positive confirming the work of Lee et al. \cite{lee2020predicting}.

\section{Pretext task group selection and weighting}
\label{sec:group}
While we now are able to predict the utility of every considered pretext task individually, the next step remains to define a clever way to combine them optimally within the same pre-training process. Hence, we introduce a method to select a group of pretext-task labels and weight their respective losses to increase or decrease their importance in the self-supervised representation. Such an optimisation of the latent representation is expected to provide better downstream performance. Our method minimises the conditional dependence between the resulting  group pretext task, entailing the prediction of a selected pretext-task label group and the downstream samples given the downstream labels.

Given a set of $k$ possible pretext-task labels $(Z_i)_{i \in [k]}$ (this notation meaning for $i$ an integer between $1$ and $k$), we seek to estimate a set of parameters $(\lambda_i)_{i \in [k]}$ weighting the loss of every pretext-task label $Z_i$ during the pre-training phase. Hence, we define a grouping pretext-task label $Z_\lambda$ as an orthogonal concatenation of $(Z_i)_{i \in [k]}$ weighted with $(\lambda_i)_{i \in [k]}$  : $Z_\lambda = (\lambda_1 Z_1, ..., \lambda_k Z_k).$ These weights $(\lambda_i)_{i \in [k]}$ will be used during the pretraining to scale the loss of every corresponding pretext task

The custom conditional HSIC computation pipeline described above is fully differentiable with respect to $(\lambda_i)_{i \in [k]}$. In the HSIC computation, the data similarity matrices $\{K_c\}_{c \in C}$ are independent of $Z$ and therefore of $\lambda$. Only the pretext-task label similarity matrices $\{L_c\}_{c \in C}$ are changed. For every downstream class $c$, $L_c$ is defined as: \vspace{-0.5em}
\begin{align}
[L_c]_{i,j} &= RBF((Z_\lambda)_i, (Z_\lambda)_j), \\
&= \exp( \frac{-1}{2\sigma^2} \sum\limits_{h=1}^{k} \lambda_h || z_{h,i} - z_{h,j} ||^2_2),\nonumber
\end{align}
where $z_{h,i}$ denotes the mean value of the $h$-th pretext-task label for the $i$-th data point in the dataset. 

 \subsection{Constraints on the weights} 
 
 The conditional-independence based utility estimator must be optimized with respect to the weighting parameters $(\lambda_i)_{i \in [k]}$ and three constraints.
 
 First, the parameters $(\lambda_i)_{i \in [k]}$ must be non negative, as they are used as weights for the corresponding losses. A negative weighting loss would lack interpretability as it could imply that the self-supervised model should ``unlearn'' the corresponding pretext task. This may be the case for adversarial learning, but we are not considering this case in this work.
 
 Second, the value of the weights must be high enough. Indeed, the presented method for estimating the conditional independence assumes that the considered pretext-task label $Z$ is not independent of $X$. It is fortunately true for speech features, as $Z$ is a function of $X$, but not always valid. For instance, with $(\lambda_i)_{i \in [k]} = 0$, the utility estimator would be zero and thus the lowest (\textit{i.e.} the best), but it would break the assumption of non independence between $Z$ and $X$, and would nullify the participation of the pretext tasks to the training loss. Furthermore, the $HSIC$ value decreases with positive decreasing values of $(\lambda_i)_{i \in [k]}$ and we thus need to ensure that the sum of the weights is significantly higher than zero to ensure that the pretext tasks are significantly considered in the pretraining loss.
 
 Finally, for a fair comparison between the weighting choices during the optimization, the sum of the weights should remain constant. In the following, the sum of the $(\lambda_i)_{i \in [k]}$ is fixed to one and the problem is summarized as follows: 
\begin{align}
\label{eq:problem}
\min_{\lambda \in \mathbb{R}^k}\quad  &HSIC( Z_\lambda, X, Y ), \;
\textrm{s.t.} \;  Z_\lambda = (\lambda_1 Z_1, ..., \lambda_k Z_k ), \\ \;
  &\lambda_i \geq 0, \:  \forall \: i \in [k], \; \nonumber
  \sum_i \lambda_i = 1. 
\end{align}
To minimize the estimator quantity while respecting the constraints, the weights used in the computation of the CI value are the softmax output of free learnable parameters $(W_i)_{i \in [k]}$. Softmax enforces the conditions while being differentiable and the problem becomes: 
\begin{align}
\label{eq:secondproblem}
\min_{W \in \mathbb{R}^k}\quad &HSIC( Z_\lambda, X, Y ), \;
\textrm{s.t.} \;  \lambda = Softmax(W) , \; \\
 & Z_\lambda = (\lambda_1 Z_1, ..., \lambda_k Z_k ).\nonumber
\end{align}
\subsection{Weights sparsity} 
 Another trait that is desirable for the weighting vector is sparsity. If a few pretext-task labels are not needed for the given downstream task, they would rather be discarded than given a low weight. First, this would save computation time including the extraction of the pretext-task labels, and their extraction and prediction during the self-supervised training process. Second, it would help with transparency to understand what features are included or not in the latent space. This sparsity property is also related to feature selection such as with LASSO \cite{grouplasso}. To ensure the sparsity of the output weighting vector, while maintaining the desired property of differentiability, we choose the sparsemax function \cite{sparsemax} to replace softmax in equation \eqref{eq:secondproblem}.  

\section{Experimental Setup} 
\label{sec:exps}

This section details the experiments validating the introduced group selection and weighting strategy, showing its addition to state-of-the-art predictive coding approaches, and testing its robustness. It, first, describes the selection and weighting processes (Section \ref{subsec:group}), the SSL models (Section \ref{subsec:models}), the downstream tasks (Section \ref{subsec:tasks}), the obtained results (Section \ref{sec:res}). Then, it shows how the technique can improve wav2vec2.0 \cite{baevski2020wav2vec} embeddings (Section \ref{sec:wav2vec}). 

\subsection{Group selection and weighting}
\label{subsec:group}

%

\begin{table}
  \caption{Candidate speech pretext-task labels and descriptions. }
  \label{tab:features}
  \centering
  \begin{tabular}{ll}
  \hline

    \textbf{Feature}      & \textbf{Description}                \\
    Loudness                    & Intensity \& approx. loudness                                       \\
    F0                   & Fundamental Frequency                                 \\
    Voicing              & Voicing Decision                     \\
    Alpha Ratio  \cite{alpha}                 & Ratio of spectrum intensity \% 1 000 Hz                               \\

    Zero Crossing Rate                & Zero crossing number per frame                 \\
    RastaSpec L1Norm              & L1 Norm of Rasta Spectrum  \cite{rasta}                                    \\
    log HNR  \cite{hnr}   & log of Harmonicity to Noise Ratio        \\
\hline
  \end{tabular}
\end{table}

In the first attempt, the same pretext tasks presented in Table \ref{tab:features} are used for the group selection and weighting experiments.  According to Figure \ref{diagram} (\textit{step 1}), we group these pretext-task labels based on their weights, \textit{i.e.} by optimising Eq. \eqref{eq:secondproblem} to obtain the different $\lambda$ values associated to each one of them. Comparative baselines follow common feature group selection strategies or ``natural'' intuitions. The first one simply bundles all the pretext-task labels together without any weighting (\textit{i.e.} $\lambda=1$ for all pretext-task labels) as proposed for PASE \cite{pascual2019learning}. As SSL group pretext-task selection is yet to be fully explored, the two other baselines come from the feature selection literature as it represents the closest field with well established techniques. The CI-based pretext-task label selection is thus compared to two well-established baselines: 

The  Maximum Relevance Minimum Redundancy (MRMR) technique \cite{mrmr} used as a baseline in this work relies on the Conditional Independence based estimator. It is close to a naive selection of the best pretext tasks according to the CI-based criterion, but  it furthermore penalizes the mutual information between the selected pretext tasks. More precisely, we select a group of $p=4$ pretext-task labels $(Z)_{i \in [p]}$ maximizing  :  
\begin{align}
Score_{MRMR}(Z) &= \frac{-1}{p} \sum_{i\in [p]} HSIC(X,Z_i|Y) \\
&- \frac{1}{\binom{p}{2}} \sum_{i<j} I(Z_i, Z_j).  \nonumber 
\end{align}

Recursive Feature Elimination (RFE) \cite{RFE} relies on a classifier that provides information concerning the importance of a given feature in the decision. This classifier is first trained with the whole set of pretext-task labels as features, and the least important feature is eliminated. The process is repeated until only 4 pretext-task labels are kept. We use the scikit-learn implementation with the C-Support Vector Classification as the classifier providing the feature importance values with the default scikit-learn hyperparameters. 

\begin{table*}[!t]
\centering
\caption{Results observed with the proposed selection strategies on the two considered downstream tasks. Word Error Rate (WER) Equal Error Rate (EER), and Accuracy (Acc) are expressed in percentage and used for LibriSpeech $100$ hours, VoxCeleb1 and IEMOCAP respectively. ASR results are given with and without Language Modeling (LM). All SSL models contain $16.3M$ neural parameters. }
\vspace*{3mm}
\scalebox{0.95}{
\begin{tabular}{@{}lcccccc@{}}
\hline
\multicolumn{1}{c}{\textbf{Models}} & \multicolumn{2}{c}{\textbf{LibriSpeech} \textit{(WER \% $\downarrow$)}}         & \multicolumn{1}{c}{\textbf{VoxCeleb1} \textit{(EER \% $\downarrow$)}}
& \multicolumn{1}{c}{\textbf{IEMOCAP} \textit{(Acc \% $\uparrow$)}}
\\ 
%

\vspace{-0.3cm}\\
 & \textit{No LM} & \textit{ LM}  & &
 \\ \hline \\
PASE+  \cite{ravanelli2020multitask}                      &25.11        & 16.62             & 11.61     &     57.86     \\ 
vq-wav2vec \cite{baevski2020vqwav2vec} & 17.71             & 12.80               & 10.38      &    58.24       \\ 
Selections                               & &              &    &       \\ 

All                             & 21.98   $\pm$ 0.36                 & 11.70  $\pm$ 0.27             & 11.90$\pm$ 0.32       & 56.4 $\pm$  1.3
\\ 

MRMR                                   & 18.94 $\pm$ 0.34                  & 10.36 $\pm$ 0.26                 & 10.56 $\pm$ 0.31    &     59.6 $\pm$1.29   \\

RFE                               & 20.02  $\pm$ 0.34                 & 11.42  $\pm$ 0.27              & 11.91   $\pm$ 0.33      &  55.8 $\pm$ 1.3  \\ 

Softmax                                 & \textbf{13.17$\pm$ 0.28}                    & \textbf{8.00 $\pm$ 0.23}               & 9.24  $\pm$ 0.29   &   60.6  $\pm$ 1.27
\\
Sparsemax                               & 17.18 $\pm$ 0.32                   & 10.41  $\pm$ 0.26              & \textbf{8.63 $\pm$ 0.27}      &   \textbf{60.8 $\pm$ 1.28 }   \\ 

\hline

\end{tabular}
}
\label{results}
\end{table*}
\subsection{Self-supervised training}
\label{subsec:models}

In the second step of Figure \ref{diagram}, the SSL model learns to predict the selected pretext-task labels. For every one of those, the loss is multiplied by the corresponding assigned weight. As for individual pretext-task testing, the network learns to reconstruct the input Mel spectrograms, and to compute $40$-dimensional Mel-Frequency Cepstral Coefficients (MFCC) feature vectors. These targets are usually kept to avoid information loss harming heavily downstream performance and are used in all our experiments. For a given weighting vector $(\lambda_i)_{i \in [k]}$, the self-supervised loss is defined as:
\begin{equation}
    L_{SSL} = MSE_{mel} + MSE_{mfcc} + \sum\nolimits_{i=1}^{k} \lambda_i \ell_1(Z_i)
\end{equation}
with $MSE$ the classic mean squared error computed for Mel spectra ($MSE_{mel}$) and MFCC ($MSE_{mfcc}$), and $\ell_1(Z)$ the $\ell_1$-loss of the pretext task related to pretext-task label $Z$. 

Prior to extending our method to state-of-the-art architectures such as wav2vec 2.0 that are particularly costly to train, we propose to first employ a PASE-like model to empirically validate the approach. The encoder and worker architectures are those described in appendix \ref{ap:details}.

The SSL model is learned on the training set of the English Common Voice dataset (version $5.1$; $700$ hours). $700$ hours of speech is a relatively small amount compared to what is generally used for state-of-the-art SSL models. However, we believe it is a sound choice as this is generally greater than what is typically available in SSL use-cases like low-resource languages. We decided to not use the LibriSpeech dataset for pre-training as it is part of our downstream evaluation protocol hence alleviating a strong bias shown in table \ref{librispeech960}. 

\subsection{Downstream tasks}
\label{subsec:tasks}
Our proposed pretext-task label selection strategy is compared with the two baselines on three different downstream tasks leading to different groups of pretext-task labels: automatic speech recognition (ASR, with LibriSpeech $100$ hours) speaker recognition (SR, with VoxCeleb 1), and emotion recognition (ER with IEMOCAP). Datasets and downstream architectures are inspired from the SUPERB benchmark \cite{superb} for self-supervised learning representations and carefully described in Appendix \ref{superbsettings}. Prior to downstream training, the SSL model are frozen to be used as a feature extractor with the new pipeline that is task-dependent. We do not use any data augmentation for a pristine comparison of the learned models.

\subsection{Extending wav2vec 2.0 to multitask SSL}
\label{sec:wav2vec}

To the best of our knowledge, multi-task speech representation learning has not been scaled to a point where it could represent a state-of-the-art alternative. Contrastive predictive coding \cite{oord2018representation} based techniques like wav2vec 2.0 \cite{baevski2020wav2vec}, on the other hand, currently trust most of the leader-boards for speech-related tasks. Recently, Sadhu et al. \cite{sadhu21_interspeech} showed that combining a consistency loss and contrastive predictive coding improves the results of the wav2vec 2.0 architectures in noisy conditions. Following this idea, we propose to further validate our selection method with an extension of wav2vec 2.0 to multitask SSL to demonstrate its scaling capabilities. Hence, the training loss is extended in a second experiment to:
\begin{equation}
    L_{SSL} = L_{W2V} + \sum\nolimits_{i=1}^{k} \lambda_i \ell_1(Z_i).
\end{equation}

We use the standard \textit{BASE} wav2vec 2.0 first described in \cite{baevski2020wav2vec} as a SSL model and train it with the same Common Voice dataset. The pre-training pipeline is implemented within SpeechBrain. The trained \textit{BASE} model has been compared to one obtained with the official Fairseq implementation from \cite{baevski2020wav2vec}, and results are strictly equivalent. The entire recipe alongside with the large set of hyperparameters needed to properly train a wav2vec 2.0 model are released under our repository and will be made available within SpeechBrain afterwards. 

We follow the SUPERB benchmark conventions \cite{superb} both at the data and downstream architecture levels. Hence, and conversely to the previous experiments, the ASR system solely optimises the CTC criterion over characters. For each of the three tasks (\textit{i.e.} ASR, SV, ER) we compare the standard \textit{BASE} Wav2vec 2.0 model with one trained following the sparsemax selection of multitask SSL. Sparsemax is chosen over softmax because it enforces the sparsity criterion and removes completely a few pretext-task labels from the training, which is one of the objectives of this work.  Another experiment is led with a ``naive'' pretext-task selection where a constant weight of $0.5$ is used across all signal-based pretext-tasks. Each wav2vec 2.0 model required $24$ NVIDIA Tesla V100 GPUs to train for $150$ epochs ($40$ hours). Finally, we also propose to compare frozen and unfrozen (\textit{i.e.} where the wav2vec 2.0 encoder is fine-tuned with the downstream task) SSL models.

\section{Experimental Results} 
\label{sec:res}
This sections details the main experiments validating the proposed approach on speech data. Table \ref{results} shows the results of the group selection methods on the three considered downstream tasks, while Table \ref{wav2vecresults} shows the impact of adding a careful selection of pretext tasks to  Wav2vec 2.0 training loss.
\subsection{Group selection results} 
Baselines detailed in Section \ref{subsec:group} are respectively referred to as ``\textit{All}'', ``\textit{RFE}'' and ``\textit{MRMR}''. First, it is clear from the results reported in Table \ref{results} that, for the considered downstream tasks, the two introduced strategies (\textit{Sparsemax} and \textit{Softmax}) perform better than the group selection baselines with a gain of $3.28$ of EER for \textit{Sparsemax} against the \textit{RFE} approach on VoxCeleb, and $8.81$ of WER with \textit{Softmax} compared to the \textit{All} baseline. Interestingly, simply bundling all the pretext-task labels together may lead to poor performance as observed on LibriSpeech with a very high $21.98\%$ of WER obtained. Hence, \textit{intuitively} building sets of labels could be harmful for the final representation. This motivates the need for a better pretext-task label selection strategy such as the one introduced in this work, as the WER dropped to $13.17\%$. As a comparison, the exact same architecture trained with Mel spectra only (\textit{i.e.} no SSL) obtains a WER of $17.3$\% without LM. Hence, our method even further decreases the WER while being only pretrained with a reasonable amount of data (\textit{i.e.} only $700$ hours compared to a few thousands for common SSL techniques \cite{baevski2020wav2vec}). As expected, introducing the joint decoding with a language model strongly decreases the WER but also introduces a bias in our comparison as probabilities are smoothed with a third-party neural model. Nevertheless, and even in this scenario, our weighting strategy outperforms all the baselines. In the context of speaker recognition, \textit{Sparsemax} beats \textit{Softmax} with an EER $0.61$ lower.  For IEMOCAP, \textit{Softmax} and \textit{Sparsemax} weighting still perform the best among all methods. To investigate how strongly improvements are correlated to the task, we took the best learned model for LibriSpeech (i.e. softmax weighting) and fine-tuned it on VoxCeleb1 and IEMOCAP. It reaches an EER of $10.55\%$ and an accuracy of $59.9\%$ respectively. While it performs better than the baselines, the difference between these results and the best performing selections shows that the weightings are indeed task-related.
\begin{table*}[t!]
\centering
\caption{Results observed training the Wav2vec2 model with and without weighted pretext tasks using the sparsemax method. ``Fr." and ``Fine." also respectively refer to Frozen and Finetuned settings. Adding selected pretext tasks improves the donwstream performance on all three considered tasks. All models contain $100M$ neural parameters. }
\vspace*{3mm}
\scalebox{0.9}{
\begin{tabular}{@{}lcccccc@{}}
\hline

\multicolumn{1}{c}{\textbf{Selections}} & \multicolumn{2}{c}{\textbf{LibriSpeech} \textit{(WER \% $\downarrow$)}}         & \multicolumn{2}{c}{\textbf{VoxCeleb1} \textit{(EER \% $\downarrow$)}} & \multicolumn{2}{c}{\textbf{IEMOCAP} \textit{(Acc \% $\uparrow$)}} 
\\ \hline \\

                                        & Fr.          & Fine.          &
                                       Fr. & Fine.  &
                                       Fr. & Fine.  \\ 
\hline
 \\
wav2vec 2.0 \textit{BASE}                             &  17.93  $\pm$ 0.33                  & 10.21 $\pm$ 0.25      &7.20 $\pm$ 0.26 &5.35  $\pm$ 0.22          & 56.6 $\pm$ 1.2     &\textbf{74.0 $\pm$ 1.16}\\ 
wav2vec 2.0 \textit{BASE}  + Naive selection                          &  17.23  $\pm$0.32                  & {10.10} $\pm$ 0.25      &6.80 $\pm$ 0.25 &\textbf{5.05  $\pm$ 0.21  }        & 57.4 $\pm$ 1.3     &73.7 $\pm$ 1.16\\ 
wav2vec 2.0 \textit{BASE} + Sparsemax                             & \textbf{16.70 $\pm$ 0.31}                   &  \textbf{9.18 $\pm$ 0.24}      &\textbf{6.57 $\pm$ 0.25}& 5.30 $\pm$ 0.22           & \textbf{59.5 $\pm$ 1.29}     &\textbf{74.0 $\pm$ 1.16}\\ \hline

\end{tabular}
}
\label{wav2vecresults}
\end{table*}
\subsection{Wav2vec 2.0 Extension results} 
Results reported in Table \ref{wav2vecresults} show that our approach improves the performance over the standard wav2vec 2.0 framework for every considered downstream task. While adding pretext tasks naively improves the final performance, the difference in performance between the naive selection and the sparsemax weighting shows the benefit of our method in getting the best downstream performance. Unsurprisingly this difference is small (though statistically significant in all but one case), as the wav2vec 2.0 BASE is already powerful and the additional workers are anyway useful.  Here, it is worth noting that the difference in performance compared to the literature mostly comes from the pre-training conditions. For instance, wav2vec 2.0 is commonly pre-trained with larger models on LibriSpeech to achieve lower WER on this dataset. 

\section{Robustness Analysis}
\label{sec:robustness}
This section explores the robustness of the method to changes in the pretraining dataset, in the audio data type and in the set of considered pretext tasks.
\subsection{Pretraining dataset robustness}
\begin{table}[t!]
\centering
\caption{Results observed retraining the Wav2vec2 model with and without weighted pretext tasks using the sparsemax method, on LibriSpeech 960. ``Fr." and ``Fine." also respectively refer to Frozen and Finetuned settings. Adding selected pretext tasks still improves the downstream performance. All models contain $100M$ neural parameters. }
\vspace*{3mm}
\scalebox{0.75}{

\begin{tabular}{@{}lcc@{}}
\hline

\multicolumn{1}{c}{\textbf{Selections}} & \multicolumn{2}{c}{\textbf{LibriSpeech} \textit{(WER \% $\downarrow$)}}         
\\ \hline \\

                                        & Fr.          & Fine.           \\ 
 \hline
 \\
wav2vec 2.0 \textit{BASE}                             &  9.88                  & 6.33       \\ 
wav2vec 2.0 \textit{BASE} + multitask SSL                             & \textbf{ 9.5 }                   &  \textbf{6.01 }     \\ \hline

\end{tabular}
}
\label{librispeech960}
\end{table}
It is common in the speech SSL literature to train on LibriSpeech 960 before fine-tuning on LibriSpeech100. As explained before, we believe that this introduces a bias due to the closeness of pretraining and finetuning data.  Studies have shown, for instance, that adding the downstream training dataset to the pretraing set of wav2vec 2.0 leads to a better downstream word error rates \cite{robustwav}. To verify this,  we train our best multitask \textit{BASE} wav2vec 2.0 architecture with the best performing pretext tasks and their weights on LibriSpeech 960. The model follows the exact same training procedure as for Table \ref{wav2vecresults}. We fine-tune the models on LibriSpeech 100 exactly as it has been done with the other models. Table \ref{librispeech960} shows the results. Two observations deserve to be noted. First, in this case also, adding a selected set of pretext tasks improves the final downstream performance in the frozen and finetuned cases. Second, as expected, the results obtained after training on Librispeech960 are better than with CommonVoice, reaching the lowest $6.01\%$  with the fine-tuned version compared to $9.18\%$ in table \ref{wav2vecresults}.
\subsection{Task and pretext tasks change robustness}
To further validate the proposed technique and test its robustness to task and data change, the following section will detail experiments led on multi-task self supervised learning for musical instrument recognition. 

\textbf{Task change : Instrument Recognition} In a first phase, the same pretext-tasks are kept and the weights are computed in a similar way. However, to be closer to the downstream task, we use AudioSet "Musical Instrument" partition for the SSL training instead of CommonVoice. The partition contains 57052 files for a total duration of 155 hours. To compute the SSL training weights, the Medley-solos-DB instrument classification dataset is used. Two reasons motivate this choice. First, the music excerpts used come from real instrument recordings as opposed to synthesized audio files from MIDI annotations. Second, every file corresponds to a single instrument played solo thus facilitating the CI estimation. We further test the representations learned in a multi-instrument setting with the OpenMIC-2018 dataset. This tests the robustness of our approach when generalising to a different downstream dataset of a slightly different task. Hence, we start by computing the pretext tasks weights corresponding to Medley-solos-DB and train the encoder using these weights. Then, it is important to note that the same encoder will be used for the two downstream tasks, i.e Medley-solos-DB and OpenMIC-2018. 

\textbf{Adding new pretext tasks candidates} In a second time, we study the impact of adding other pretext tasks to the pool of candidates. To investigate this, we select three additional new candidates: mean spectral centroid, mean spectral kurtosis and Hammarberg Index. After adding these features, the sparsemax weighting is computed as in \ref{eq:secondproblem}. It is interesting to note first, that two of the three features are not selected for pretraining (even when added individually), thus not changing the weighted selection.  Mean spectral centroid is the only feature selected for pretraining, lowering the weights of the other selected tasks. We will refer to this experiment as sparsemax+ in the results table (Table \ref{musicresults}).

\textbf{Are MFCCs essential?} A final change is considered. Following the works of Ravanelli et al. \cite{ravanelli2020multitask} and their ablation studies, all the experiments considered MFCCs as one of workers with a fixed unit weight. Furthermore, when studying ablations, MFCC shows the highest contribution.  While the Mel spectrogram reconstruction is needed to avoid any information loss, the MFCC worker can be weighted as well or replaced with other common time-frequency based representations. To explore this choice and its impact, we select four candidates including MFCC, with SpectralFlatnessPerBand (SFPB) \cite{herre2001robust}, Octave band signal intensity (OBSI) \cite{essid2005classification}, and Chroma. These features are computed using the Yaafe toolkit \cite{yaafe}. 

The kernel used for these features is the same used for the speech samples, i.e gaussian downsampling followed by the Frobenius inner product. As in the previous paragraph, we compute a sparsemax based selection on these four candidates along with the initial best selection of weighted pretext-tasks (without the Spectral Centroid addition). The pretraining loss therefore becomes: 
\begin{equation}
    L_{SSL} = MSE_{mel} + \sum\nolimits_{i=1}^{l} \mu_i \ell_1(S_i) + \sum\nolimits_{i=1}^{k} \lambda_i \ell_1(Z_i)
\end{equation}
with $(S_i)_{i \in [l]}$ the spectral representations, and $\sum\nolimits_{i=1}^{l} \mu_i =1$. This experiment will be refered to as "spectral+" in the results.

\textbf{ Downstream datasets and architectures} Medley-solos-DB contains 21572 3-second audioclips distributed among 11 classes. OpenMIC-2018 contains $20,000$ musical samples with partial instruments annotation for $20$ instruments. Although not every file is labeled for every instrument, each class has at least $500$ confirmed positives, and at least $1,500$ confirmed positives or negatives. We adopt for downstream finetuning, an X-vector like architecture, similar to the one used for VoxCeleb1 for Medley-solos-DB. For OpenMIC-2018, we use the official baseline technique relying on a random forest classifier for every considered instrument. 

The same grouping techniques presented in the previous section are compared here. Results on the two datasets are shown in table \ref{musicresults}. We highlight the best results with the standard pool of pretext tasks, and  the best score after the additions.  Accuracy on the test set is computed for Medley-solos-DB while mean F1 Score is shown for OpenMIC following the SSL literature for music classification \cite{wu2021multitask}. The results follow those on speech processing tasks both for Medley-solos and OpenMIC. This confirms that the method presented generalizes to other types of data, another pretraining dataset and to downstream tasks that are similar to the one used for weights computing. OpenMIC best selection results are 3 points higher than the selection done in \cite{wu2021multitask}.  Running the experiment on instrument classification with the three additional pretext tasks in the pool of candidates leads to an even better classification reaching $76.1$\%. This confirms literature findings on the importance of Spectral Centroids in timbre classification. The same model performs better on OpenMIC reaching $66.0$ mean F1-score. This suggests that the selection technique is not harmed by adding irrelevant features, while adding relevant ones can improve the final results. 
Finally, replacing the MFCC tasks by a weighted combination of spectral representations achieves a better result than the Sparsemax selection on Medley-solos-DB with a score of $74.6$\%. It also reaches the best result on OpenMic with $67.7$ mean F1-Score.
\begin{table}[!t]
\centering
\caption{Results observed with the proposed selection strategies on the two considered downstream instrument recognition tasks. Accuracy on the test set is computed for Medley-solos-DB while mean F1 Score is shown for OpenMIC. Higher is better for both.}
\vspace*{3mm}
\scalebox{0.8}{
\begin{tabular}{@{}lccc@{}}
\hline

\multicolumn{1}{c}{\textbf{Models}} & \multicolumn{1}{c}{\textbf{Medley-solos} \textit{(Acc\% $\uparrow$)}}         & \multicolumn{1}{c}{\textbf{OpenMIC-2018} \textit{(mean-F1   $\uparrow$)}}

 \\ \hline \\PASE+ \cite{ravanelli2020multitask}                  & None        & 64.1                
 \\ \hline \\
Selections                               & &             \\ 

All                             & 66.2  $\pm$ 0.83                 & 62.89          
\\ 

MRMR                                   & 62.3 $\pm$ 0.85                  & 64.23              \\

RFE                               & 64.6  $\pm$ 0.84                 & 62.80             \\ 

Softmax                                 & \textbf{73.5$\pm$ 0.78}                    & 65.06         
\\
Sparsemax                               & 72.6 $\pm$ 0.79                   & \textbf{65.39}       \\ 
Sparsemax+                                 & \textbf{76.1$\pm$ 0.76}                    & 66.0         
\\
Spectral+                              & 74.6 $\pm$  0.77                   & \textbf{67.7}         
\\
\hline
\end{tabular}
}
\label{musicresults}
\end{table}

 \begin{figure*}[!t]
  \centering
  \includegraphics[width=0.8\linewidth, scale=0.25]{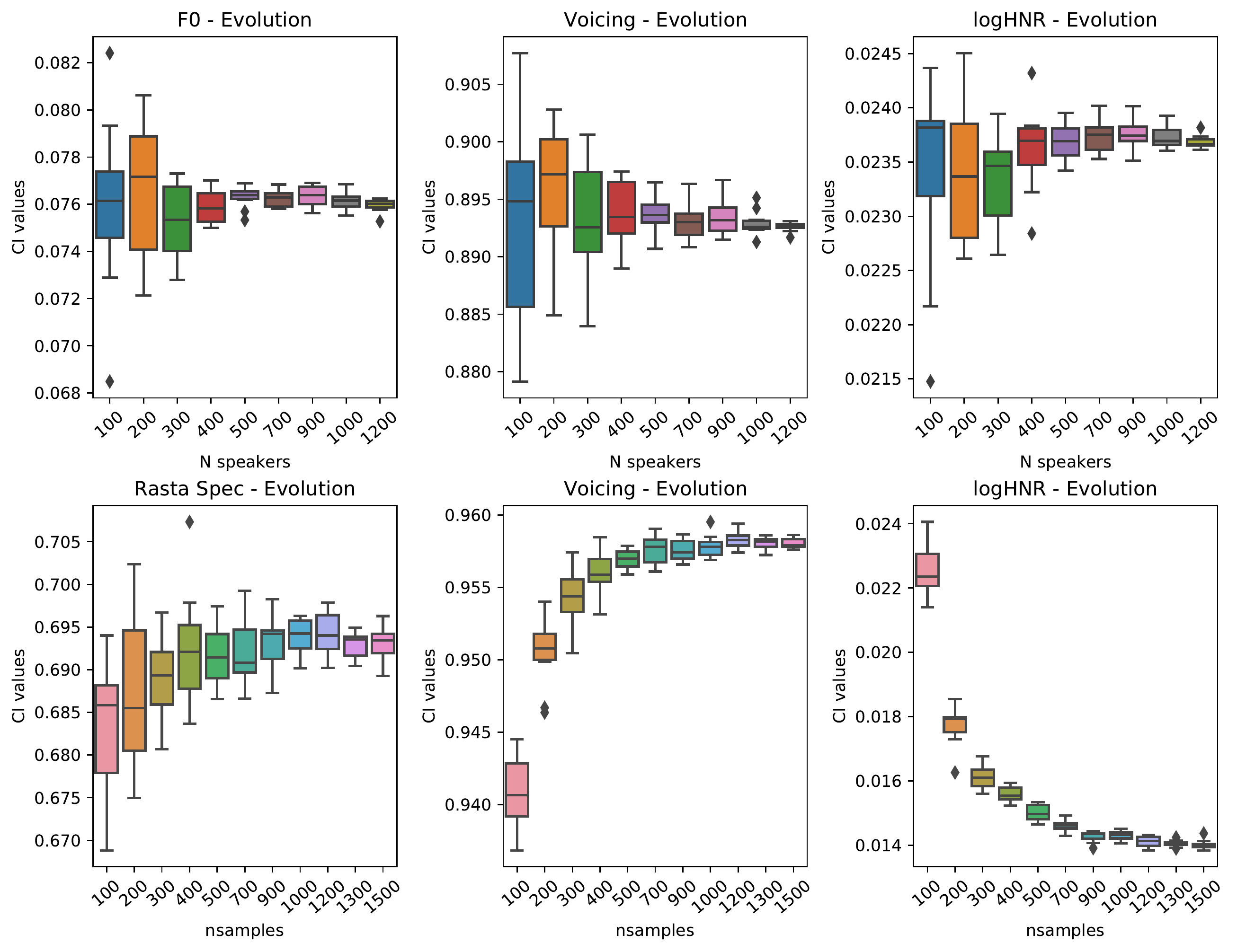}
  \caption{ Evolution of the CI estimation with different numbers of considered speakers for VoxCeleb (First row of plots) and number of samples for Medley (Second row of plots), for three pretext tasks : F0, Voicing and logHNR, Rasta Spech. We can see that the values obtained with 20 speakers and 100 samples per class, while logically exhibiting more variance, are already close to the final values for every pretext task.} 
  \label{boxplots}
\end{figure*}
 \begin{figure*}[!t]
  \centering
  \includegraphics[width=0.6\linewidth, scale=0.1]{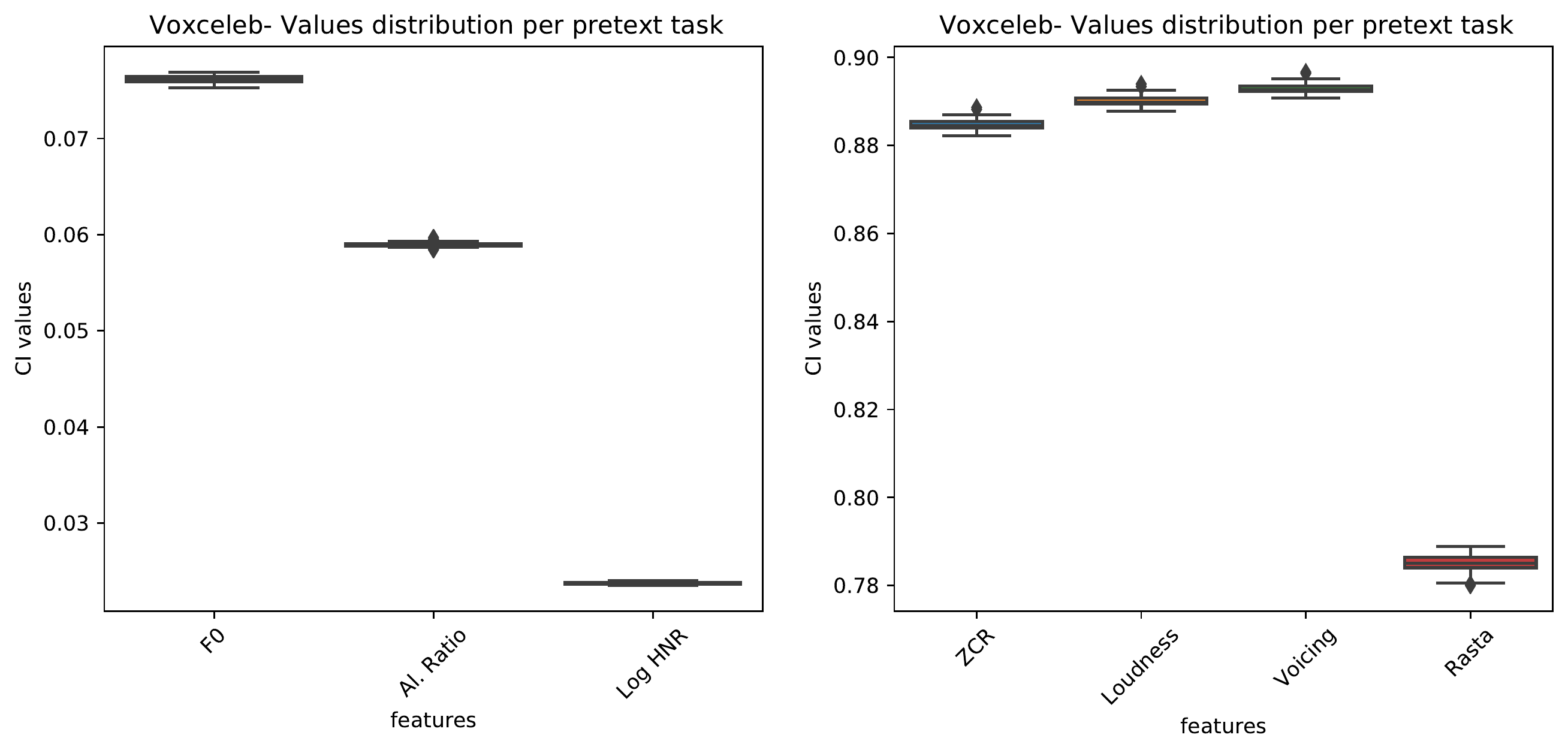}
  \caption{ Boxplots of the CI values for every pretext tasks, when more than 200 speakers are considered. Voicing and Loudness are slightly overlapping, but otherwise, the values are separable. We divide the pretext-tasks in two groups according to their CI values for a better visualisation of the results.}
  \label{distribution}
\end{figure*}
\section{Computational efficiency}
 Efficiency is one of the key motivations of this work, and the gain in time observed with our approach is considerable. The $K$ and $L$ matrices used for the CI estimate are only computed for the downstream datasets. Two limitations related to the size of the downstream dataset may be faced using our technique. First, very small downstream datasets could not be sufficient for a good estimate of the conditional independence. Second, very large downstream datasets may render the CI estimation intractable as the matrices involved get larger. It is important to note here that the computations needed in order to obtain the weights are performed on the downstream dataset, and not on the pretraining unlabeled one. This means that enlarging the unlabeled dataset does not lead to heavier computations.
 
 This section shows experimentally on VoxCeleb1 and Medley-solos-DB that our technique is robust to these two situations. First, we show by taking small subsets of VoxCeleb1 and Medley-solos-DB, that in case of downstream data scarcity, the CI estimations obtained with our method are close to the final estimations, and the ranking of the pretext tasks is not altered even when we take only 200 speakers among the $1 \:251$ in VC. Second, as one of the main  motivations of this work is the reduction of the computation needed to get the best selection of pretext-tasks in self-supervised learning settings, we show that the CI estimation converges quickly with a growing number of speakers considered, and is thus resilient to sampling. Considering one pretext task at a time, we consider subsets of VoxCeleb1 using a growing number of considered speakers ($total=1 \:251$), and subsets of Medley-solos-DB using an increasing number of samples per considered instrument class. For each of these considered numbers, we run $10$ experiments with sampled speakers and music excerpts. We get the CI estimation for every subset and plot the boxplot of the obtained values. Results are shown in Fig. \ref{boxplots}. We can see that using only $20$ speakers exhibits results that are already close to those with $1 \: 000$ speakers, and the results using 100 audio files per class are close to those with 1500 points per class. Furthermore, we plot the boxplots of CI values obtained using more than $200$ speakers to show the separability between the considered features in Fig. \ref{distribution}. While values for Voicing and Loudness are slightly overlapping, all the other pretext tasks are already separated and rankable using only $200$ random speakers among the whole dataset.
 
 Training the model with a random selection of pretext tasks takes about 30 hours on a single V100 GPU, for the basic model and 40 hours on 24 GPUs for the wav2vec2.0 enhanced one. Finding, through an empirical random search, the best combination of pretext tasks takes a number of  experiments that is exponential in the number of pretext tasks considered in the initial set. In contrast, and using as done in this paper, 50 random speakers of VoxCeleb, the whole computation of the optimal weights is performed in a few hours (6 approximatively) when parallelized on 20 CPUs. This runtime is divided into extracting the pretext-task labels, running the gaussian downsampling of the speech samples (the longest part, as it involves computing the Mel spectrograms before downsampling them), computing the similarity matrices and finally optimizing the HSIC quantity according to the weights. The same durations are observed with LibriSpeech, where the number of points is higher in our experiments, but the speech segments are shorted since we cut them at the word level.

\section{Acknowledgements}
This work has benefited from funding from l'Agence de l'Innovation de Défense, and was performed using HPC resources from GENCI-IDRIS (Grant 2021-AD011012801R1 and AD011012633).

\section{Conclusion}
 In this work, we introduce a method to quickly and simply combine pretext-task labels into a useful pretext task for multitask self-supervised learning settings. Our approach allows for an optimal selection of pretext-task labels following a cheap optimisation process drastically decreasing the time and compute needed to design the best performing multitask SSL model. Our method is validated on three speech-related downstream tasks and on musical instrument recognition and outperforms common pretext-task label selection strategies when combined with simple and state-of-the-art SSL models. This opens a range of possibilities for finding and selecting new pretext tasks in self-supervised learning for speech, audio or other types of data. 
\bibliographystyle{IEEEtran}
\bibliography{mybib}

{\appendix[Training and architectures details]
\label{ap:details}

All the considered audio files are sampled at 16kHz. We feed the SSL models with $80$-band Mel spectrograms, with $25$ms windows and $10$ms stride. To every Mel band corresponds a learned vector of size $256$ obtained at the output of the SSL model. So if the input spectrogram is of size ($N=80$) with $N$ the number of frames, the representation fed to the downstream pipeline is of size ($N=256$). All models including SSL and downstream ones are developed with SpeechBrain \cite{speechbrain}.

\subsection{Pretraining of the PASE-like SSL encoder.} 
The encoder is a succession of 2D CNN layers, LSTM layers and a final dense network. This representation is then fed to one dense layer that predict the selected pretext task labels. There are $3$ successive CNN blocks containing each $2$ CNN layers  with kernel size $(3,3)$ and $128$, $200$ and $256$ channels for each block respectively. No time pooling is performed in order to preserve the input sequence length. $5$ bidirectional LSTM layers of size $256$ are then stacked. Finally, a MLP with one hidden layer with $256$ neurons. The LeakyReLU activation is used across all the layers except for the LSTM. We use a dropout rate of $0.15$ during the training. The AdaDelta optimizer is used to update the weights with an initial learning rate of $1.0$, $\rho=0.8$ and $\epsilon=10^{-8}$. For every experiment, the SSL model is trained for $10$ epochs ( leading to the convergence of the  
validation loss).

\textbf{Speaker recognition details.} 
 VoxCeleb1 \cite{Nagrani_2017} is used for the speaker recognition task. The training set contains $148,642$ utterances from $1,251$ different speakers. To compute the conditional independence estimates while limiting the computational load, we restricted ourselves to the utterances of $50$ different speakers (the detailed list is given in the released repository  \url{https://github.com/salah-zaiem/Multitask-pretext-task-selection}). A standard xvector model \cite{snyder} is trained following the available VoxCeleb SpeechBrain recipe. The extracted speaker embeddings are tested on the enrol and test splits using PLDA \cite{plda} as a similarity metric. Performance is reported in terms of equal error rate (EER). While architecture details are given in appendix \ref{ap:details}, it is worth noticing that the whole pipeline is fully integrated to Speechbrain and can thus easily be extended. 
 
We train an embedding model (XVector) until the validation loss converges, on top of the self supervised representations using $5$ successive layers of time-delay neural networks (TDNN) \cite{Peddinti2015ATD}. The number of channels is $(512,512,512,512, 1500)$, with kernel sizes of $(5,3,3,1,1)$ and dilations of $(1,2,3,1,1)$. The architecture is inspired by successful works on embeddings for speaker recognition \cite{tdnnsnyder}. The learned embeddings are therefore used on a list of pairs of samples to predict whether they are from the same speaker or not. The details of the recipe can be found in the given GitHub repository. We train every embedding model on 10 epochs with an Adam Optimizer starting with a learning rate of 0.001 decaying linearly to 0.0001.

\textbf{Speech recognition details.} ASR is conducted with the $100$-hour clean subset of the LibriSpeech dataset \cite{librispeech} to simulate the low-resource scenario commonly encountered with SSL settings. CI estimations are obtained with word-level alignments from the \textit{Montreal Forced Aligner} \cite{mfa}. The ASR pipeline follows the LibriSpeech recipe of SpeechBrain \cite{speechbrain} and therefore contains a CRDNN encoder (\textit{i.e.} CNN, RNN, DNN) trained jointly with CTC \cite{graves2012connectionist} and attention \cite{L_scher_2019} (details in appendix \ref{ap:details}). The decoding process is based on beam-search with and without shallow fusion with a pretrained recurrent language model that is publicly available and obtained from SpeechBrain.\footnote{ \url{https://huggingface.co/speechbrain/asr-crdnn-rnnlm-librispeech}} Performance is expressed in word error rate.

The CRDNN starts with three CNN blocks composed each with two 2D CNN layers, layer-normalisation and $(2,2)$ maxpooling along the frequency dimension. The filter dimensions for each block are $64, 100, 100$. Then, maxpooling of $4$ is applied on the time dimension to reduce the sequence length before being fed to the RNN. The latter is made of $5$ bidirectional LSTM layers of $1,024$ neurons. Finally two dense layers are connected (with batch-normalisation in between). The LeakyReLU activation function is used across all the layers except for the LSTM. A dropout rate of 0.15 is employed with the encoder. The CTC decoder is a simple dense linear layer of size equal to the vocabulary. The vocabulary is obtained with byte pair encoding or sub-words units (BPE) and is of size $1,000$. The attentional decoder is a one-layered location-aware GRU ($1,024$ neurons). Then, a beam search of depth $60$ is applied to obtain the output transcripts. The model is trained for $30$ epochs. The learning rate ($1.0$) is multiplied with a factor of $0.8$ every time the validation loss is not decreasing to ensure an optimal convergence of all the models.  

\subsection{SUPERB settings} \label{superbsettings}
SUPERB \cite{superb} is a recent benchmark for self-supervised representations of speech data. We use this benchmark for our experiments in combining wav2vec with our selected pretext tasks. We detail here the downstream models as detailed in the benchmark paper :

\textbf{Emotion Recognition.} IEMOCAP \cite{iemocap} is used for the Emotion Recognition (ER) task. 4 classes are considered (neutral, happy, sad, angry), and only the audio data is used. The learned representations are mean-pooled then fed to a final linear classifier to compute a cross-entropy loss. We cross-validate on five folds of the standard splits. The result shown is the average of the five attempts. The evaluation metric is accuracy (ACC).

\textbf{Automatic Speech Recognition} For ASR, the decoder is a vanilla 2-layer 1024-unit BLSTM fed with our self-supervised representations and optimized by CTC loss on characters. We use the same language model for decoding as in the first experiments. LibriSpeech Clean-100 only is used for downstream training.

\textbf{Speaker Recognition} The model and the dataset splits used in the first experiment correspond to the SUPERB ones, so we kept the same settings. The results are therefore comparable.
}
\section{Pretext-task labels' interactions.} 
\begin{figure*}[!t]
  \centering
  \includegraphics[width=0.50\linewidth, scale=0.065]{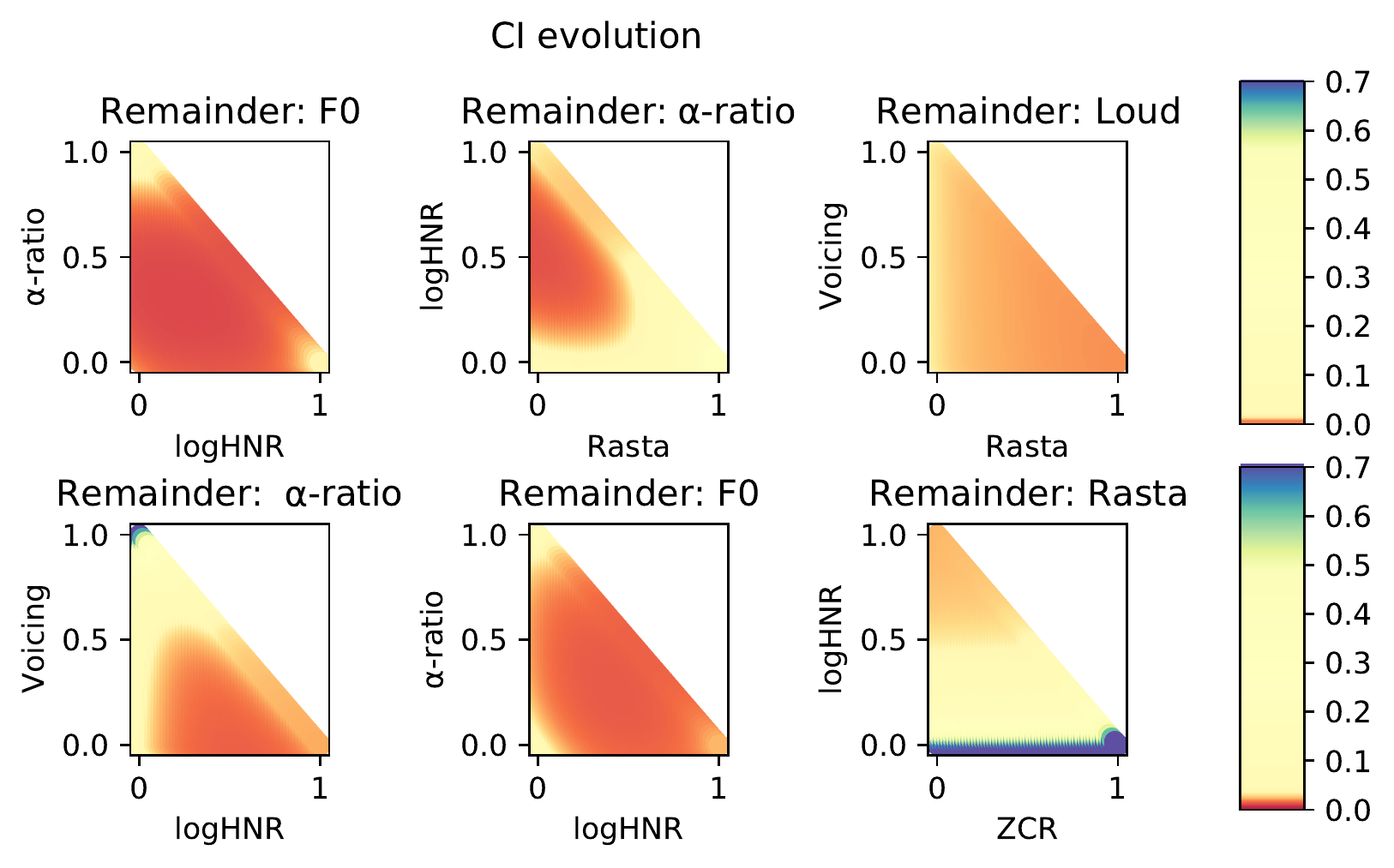}
  \caption{ CI-Based utility estimator as a function of the weighting for groups of three pretext-task labels. Top line is for Librispeech, while the bottom one is for VoxCeleb. Three pretext-task labels are presented on every plot, one on the $x$-axis, one on the $y$-axis and one that is equal to $1-x-y$ (hence being called the remainder) and whose name is on the title. Every point in the triangle corresponds to a pretext task that is the weighted combination of the three considered pretext-task labels. For instance, in the top left corner, the point $(0.5,0.3)$ correspond to the CI value of a pretext task weighting logHNR with $0.5$, $\alpha$-ratio with $0.3$ and F0 with $0.2$.}   
  \label{triangles}
\end{figure*}

To understand the interactions between pretext-task labels, studying the evolution of the CI estimate as a function of the weights shows which pretext-task labels seem interchangeable, which ones are complementary and which ones seem only harmful to the considered downstream task. Figure \ref{triangles} shows the CI estimates for weighted combinations of groups of three pretext-task labels. As the weights sum up to one, two pretext tasks' values are shown on the $x$ and $y$ axes, while the value of the remaining one, whose name is in the title, is equal to $1-x-y$. For instance, at the origin point $(0,0)$, only the third pretext-task label is selected with a weight equal to one, while its weight is equal to zero on the hypotenuse of the right triangle. Figure \ref{triangles} illustrates that the relationship leading to a lower CI-based utility estimator is not always straightforward. For instance, if we consider the second plot on the second row (\textit{i.e. $\alpha$-ratio, F0, logHNR}), we can see that selecting only one element is always worse than selecting a weighted concatenation, because the areas around the origin and the points $(1,0)$ and $(0,1)$ are brighter than the central area. 
\begin{table}
\centering
\caption{Weights for every pretext task in every experiment. When the technique only outputs a selection of the pretext tasks, $1$ is assigned as a weight for the selected tasks and zero for the non selected. This table confirms the sparsity induced by the Sparsemax function.}
\vspace*{3mm}
\scalebox{0.6}{
\begin{tabular}{@{}lccccccccc@{}}

\textbf{Selection} &$\alpha$-zero & F0 & Loudness & Spec Rasta & ZCR & log HNR & Voicing
\\ \hline \\

                             All            & 1 &1 &1 & 1& 1& 1 & 1  \\ 

                             VC RFE            & 1 &1 &0 & 0& 1& 0 & 1  \\ 
                             VC MRMR            & 1 &0 &0 & 1& 0& 1 & 0  \\ 
                             VC Sparsemax            & 0.28 & 0.26 &0 & 0& 0& 0.45 & 0  \\ 
                             VC Softmax            & 0.27 &0.11 &0.18 &0.04 & 0.06& 0.31 & 0.03  \\ 
                             Libri RFE            & 1 & 0 &0 & 0& 1& 1 & 1  \\ 
                             Libri MRMR            & 0 &1 & 0& 1& 0& 1 & 1  \\ 
                              Libri Sparsemax             & 0.30 &0.37 &0 & 0.06& 0& 0.27 & 0  \\ 
                             Libri Softmax            & 0.28 &0.47 &0.07 & 0.04& 0.02& 0.08 & 0.04  \\ 
                                                          IEMO RFE  & 0 &0 &1 & 1& 1& 1 & 0 \\
                             IEMO MRMR   & 0 &1 &0 & 0& 1& 1 &1\\

                             IEMO Spa            & 0.16 &0.22 &0 & 0.14& 0.12& 0.17 & 0.19  \\ 
                             IEMO Soft   & 0.29 &0.32 &0.06 & 0.24& 0.03& 0.02 & 0.03 \\

\end{tabular}}
\label{weighting}
\end{table}
\section{Biography Section}
\begin{IEEEbiography}[{\includegraphics[width=1in,height=1.25in,clip,keepaspectratio]{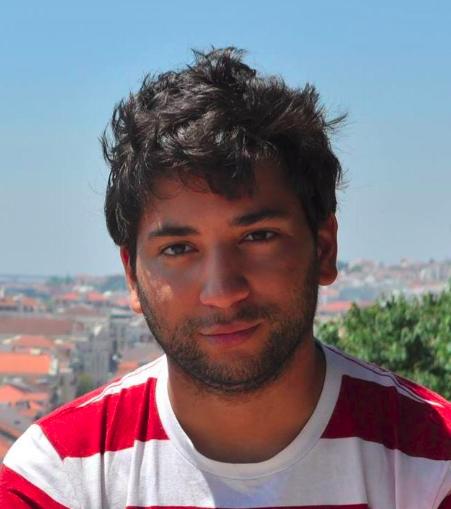}}]{Salah Zaiem}
is currently a PhD student at Telecom Paris under the supervision of Slim Essid and Titouan Parcollet. Before that, he graduated from Ecole Polytechnique in Applied Mathematics and Computer Science. He also holds a masters degree from ENS Paris-Saclay (MVA Program) in Machine Learning. His current research focuses on understanding and motivating the choices in Self-supervised learning pipelines for speech.
\end{IEEEbiography}
\begin{IEEEbiography}[{\includegraphics[width=1in,height=1.25in,clip,keepaspectratio]{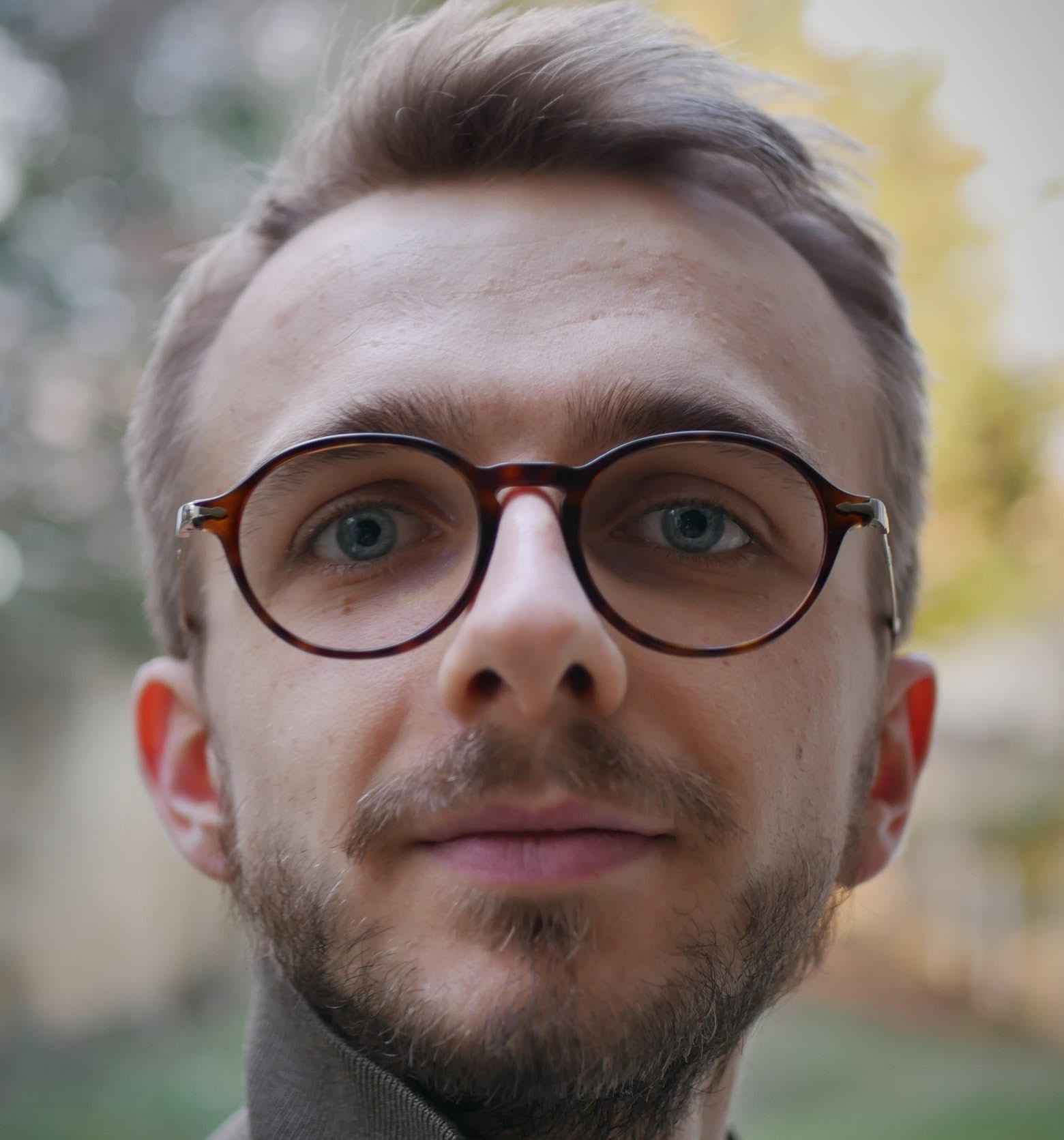}}]{Titouan Parcollet}
 is an associate professor in computer science at the Laboratoire Informatique d’Avignon (LIA), from Avignon University (FR) and an adjunct researcher at the Cambridge Machine Learning Systems Lab from the University of Cambridge (UK). Previously, he was a senior research associate at the University of Oxford (UK) within the Oxford Machine Learning Systems group. He received his PhD in computer science from the University of Avignon (France) and in partnership with Orkis focusing on quaternion neural networks, automatic speech recognition, and representation learning. His current work involves efficient speech recognition, federated learning and self-supervised learning. He is also currently collaborating with the Mila-Quebec AI institute as a co-leader of the SpeechBrain project.

\end{IEEEbiography}
\begin{IEEEbiography}[{\includegraphics[width=1in,height=1.25in,clip,keepaspectratio]{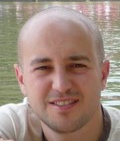}}]{Slim Essid}
is a Full Professor at Institut Polytechnique de Paris, Télécom Paris, and the coordinator of the
Audio Data Analysis and Signal Processing (ADASP) group. His research interests are in machine
listening and more generally machine learning and signal processing for temporal data analysis.
He received the Ph.D. degree from the Université Pierre et Marie Curie (UPMC), in 2005; and the
habilitation (HDR) degree from UPMC in 2015.
He has been involved in various collaborative French and European research projects and has
published over 130 peer-reviewed conference and journal papers with more than 50 distinct
co-authors. On a regular basis he serves as a reviewer for various machine learning, signal
processing, audio and multimedia conferences and journals, for instance various IEEE transactions,
and as an expert for research funding agencies.
\end{IEEEbiography}

\begin{IEEEbiography}[{\includegraphics[width=1in,height=1.25in,clip,keepaspectratio]{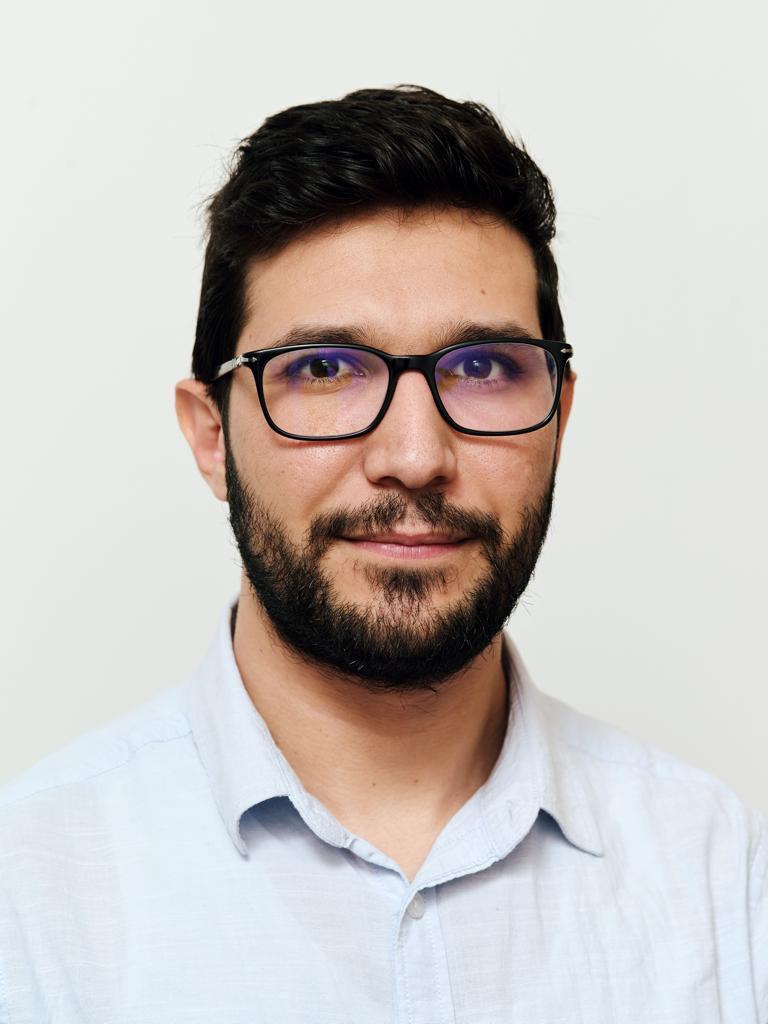}}]{Abdelwahab Heba}
 has received his Phd from Institut de Recherche en Informatique de Toulouse (France) in 2021, His research mainly focusrs on Speech Recognition and Spoken Language Understanding, He has several industrial expertise and collaborations in speech industry. He is also an active collaborator with the Mila-Quebec AI institute as a core member of the SpeechBrain project.
\end{IEEEbiography}

\vspace{11pt}

\vfill
\end{document}